\documentclass[twocolumn,english,pra,showpacs,superscriptaddress,floatfix,twocolumn,preprints]{revtex4}
\usepackage{mathptmx}
\usepackage[T1]{fontenc}
\usepackage[latin9]{inputenc}
\usepackage{amsmath}
\usepackage{graphicx}

\makeatletter
\@ifundefined{textcolor}{}
{%
 \definecolor{BLACK}{gray}{0}
 \definecolor{WHITE}{gray}{1}
 \definecolor{RED}{rgb}{1,0,0}
 \definecolor{GREEN}{rgb}{0,1,0}
 \definecolor{BLUE}{rgb}{0,0,1}
 \definecolor{CYAN}{cmyk}{1,0,0,0}
 \definecolor{MAGENTA}{cmyk}{0,1,0,0}
 \definecolor{YELLOW}{cmyk}{0,0,1,0}
 }

\makeatother

\usepackage{babel}

\begin{document}

\title{All-optical diode action in asymmetric nonlinear photonic multilayers
\\
with perfect transmission resonances}

\author{Sergei V. Zhukovsky}

\affiliation{Department of Physics and Institute for Optical Sciences, University
of Toronto, 60 St.~George Street, Toronto, Ontario M5S 1A7, Canada.}

\email{szhukov@physics.utoronto.ca}

\author{Andrey G. Smirnov}

\affiliation{B.~I.~Stepanov Institute of Physics, National Scademy of Sciences
of Belarus, Pr.~Nezalezhnasti 68, 220072 Minsk, Belarus.}

\email{a.smirnov@dragon.bas-net.by}

\pacs{42.65.Pc, 78.67.Pt, 42.25.Bs.}
\begin{abstract}
Light propagation in asymmetric Kerr-nonlinear multilayers with perfect
transmission resonances is theoretically investigated. It is found
that hybrid Fabry-Pérot/photonic-crystal structures of the type $(\text{BA})^{k}(\text{AB})^{k}(\text{AABB})^{m}$
exhibit both pronounced unidirectionality (due to strong spatial asymmetry
of the resonant mode) and high transmission (due to the existence
of a perfect transmission resonance). This results in nonlinear optical
diode action with low reflection losses without need for a pumping
beam or input pulse modulation. By slightly perturbing the perfect
transmission resonance condition, the operating regime of the optical
diode can be tuned, with a trade-off between minimizing the reflection
losses and maximizing the frequency bandwidth where unidirectional
transmission exists. Optical diode action is demonstrated in direct
numerical simulation, showing >92\% transmittance in one direction
and about 22\% in the other. The effect of perfect transmission resonance
restoration induced by nonlinearity was observed analytically and
numerically. The proposed geometry is shown to have advantages over
previously reported designs based on photonic quasicrystals. 
\end{abstract}
\maketitle

\section{Introduction\label{sec:INTRODUCTION}}

\global\long\def\sa{\text{\ensuremath{S_{1}}}}
\global\long\def\sb{\text{\ensuremath{S_{2}}}}
\global\long\def\sab{\text{\ensuremath{\bar{S}_{1}}}}
\global\long\def\sboth{\text{\text{\ensuremath{S_{1,2}}}}}
\global\long\def\mwcm{\text{MW}/\text{cm}^{2}}
An optical device capable of unidirectional light transmission (an
\emph{optical diode}) is a direct analogy to an electronic diode and
has a great application potential in optical communication and information
processing. To this end, a variety of effects can be used, including
magnetooptical \cite{diodeFaraday00,diodeFaradayK07,diodeFaradayA07},
liquid-crystal \cite{diodeLC05}, nonlinear \cite{diodeNLScalora94,diodeNLScalora95,diodeNLAssanto99,diodeNLAssanto01,FabioJAP08},
or gyroanisotropic \cite{diodeChiral07} elements in photonic microstructures.
Also, several new concepts have been proposed recently such as indirect
photonic interband transitions \cite{diodeFanInterband09}, planar
chiral \cite{diodePCMPlum09,diodePCMOurOL09} or negative-refraction
\cite{diodeLHM05} metamaterials. Among all these approaches, nonlinear
one-dimensional (1D) photonic multilayers deserve special attention
because of their extreme simplicity in fabrication and modeling. This
allows to explore the optical properties of non-periodic structures
systematically and in a wide range of geometries (see, e.g., \cite{our2008}).
In addition, a potential for versatile ultrafast pulse control by
multilayer structures only a few microns thick has been shown in recent
experiments \cite{thinfilmDefect,thinfilmFibonacci}. 

The operation of a nonlinear photonic multilayer-based optical diode
relies on two principles. First, the structure should be strongly
\emph{nonlinear}. A spectral resonance is usually employed to enhance
the nonlinear interaction, also providing the structure with a highly
transmissive state which is very sensitive to parameter variations.
Second, the structure must be spatially \emph{asymmetric} so that
the resonant mode couples more strongly to the wave impinging from
one side (e.g., from the left) than from the other side. This causes
a direction-dependent refractive index change due to the Kerr nonlinearity.
So, the leftward-impinging wave can induce a sufficient shift in the
resonant frequency $\omega_{\text{res}}$ so that it matches the incident
wave frequency $\omega_{0}$, resulting in a high left-to-right transmission,
while the rightward-impinging wave causes a weaker shift of $\omega_{\text{res}}$
so that the $\omega_{\text{res}}\neq\omega_{0}$, resulting in a lower
right-to-left transmission. Several design ideas have been proposed,
achieving spatial asymmetry using graded refractive index \cite{diodeNLScalora94,diodeNLScalora95}
or asymmetrically placed defects \cite{diodeNLAssanto99,diodeNLAssanto01,diodeLHM05}
in a 1D photonic crystal (PhC), or, more recently, using inherently
asymmetric eigenstates in photonic quasicrystals \cite{FabioJAP08}.
Generalizations of the approach beyond the 1D multilayer geometry
have also been reported, based on quasi-1D coupled-resonator \cite{LinLanChinese05,LinLanJOSA06}
or 2D \cite{MingaleevJOSA02,LinLanOE08} PhC waveguides.

Nevertheless, for a vast majority of photonic structures, the design
principles for an optical diode seem to conflict on one important
point. To increase the structure's sensitivity to the direction of
incidence, one needs to increase the spatial asymmetry of the structure.
However, to increase the maximum transmission at a resonance peak,
the structure should remain symmetric. It was once believed that only
symmetric multilayers can exhibit theoretically perfect transmission
resonances (PTRs) \cite{symHuang1,symHuang2,symMauriz4} and that
PTRs are explicitly related to mirror symmetry \cite{symHuang3gen}.
Further studies revealed that perfect transmission is rare but possible
in asymmetric multilayers. Examples based on periodic \cite{asymmHuang},
Fibonacci \cite{Nava09}, and Thue-Morse \cite{FabioTaconaPNFA10}
geometry were given. In a series of recent works, Grigoriev and Biancalana
\cite{FabioTaconaPNFA10,FabioNewJP10} report optical diode action
based on PTRs in Thue-Morse multilayers. However, such structures
typically need to consist of many layers due to relatively weak spatial
asymmetry and field enhancement, increasing their sensitivity to the
detrimental effects of material absorption and fabrication imperfections.
Besides, quasiperiodic structures possess a very rich variety of resonant
modes, not all of which lend themselves to straightforward studies.
Although there has been definite progress in analytical treatment
of nonlinear resonances in Thue-Morse multilayers \cite{FabioNewJP10},
the overall picture remains complicated. 

In our recent work \cite{ourPRA10}, it was confirmed that mirror
symmetry is sufficient but by no means necessary to design multilayers
with PTRs. It was shown that structures combining perfect transmission
and highly asymmetric light localization at the resonant modes can
readily be achieved in a simple Fabry-Pérot/PhC (FPPC) geometry of
the type $(\text{BA})^{k}(\text{AB})^{k}(\text{AABB})^{m}$. Moreover,
localization strength and asymmetry can be straightforwardly and independently
controlled by changing $k$ and $m$, respectively (see Fig.~\ref{fig:FIG1-intro}).
It was envisioned that such structures would bring about an improved
design of a nonlinear optical diode. However, an explicit investigation
of this design in the nonlinear regime has not been performed.

In this paper,\textbf{ }we investigate the nonlinear optical response
of FPPC multilayers from the point of view of optical diode design.
It is shown that FPPC structures are suitable for high-transmission
and high-contrast unidirectional operation, resulting from combined
contribution of PTRs and spatial asymmetry. A tradeoff between the
strength of asymmetric response and the amount of reflection losses
is established, so the best unidirectional operation is reported in
a design where the PTR condition is slightly perturbed. Moreover,
the effect of PTR restoration was observed where nonlinearity is seen
to increase the maximum transmittance in a frustrated PTR back to
unity. The optimized design exhibits optical diode action in the same
input intensity range (around 10 $\mwcm$) but with less than half
as many layers as a Thue-Morse structure reported in Refs.~\cite{FabioNewJP10,FabioTaconaPNFA10},
owing to an increased localization strength in FPPC multilayers. By
means of direct time-domain numerical simulations, optical diode action
in the passive regime was demonstrated with more than 92\% transmission.
In the same parameter range, only pump-assisted or active operation
was reported previously \cite{LinLanChinese05,FabioNewJP10,FabioTaconaPNFA10}. 

The paper is organized as follows. In Section~\ref{sec:THEORY},
the prerequisites needed for PTR formation in FPPC structures are
briefly reviewed. In Section~\ref{sec:PARAMETERS}, the nonlinear
response of these structures is analyzed by using nonlinear transfer
matrix method. The role of asymmetry and PTRs in optical diode action
is identified. It is found that perturbation of the PTR condition
leads to a tradeoff between reflection losses and unidirectionality,
and that a proper structure design will cause PTR restoration in the
nonlinear regime. In Section~\ref{sec:FDTD}, direct numerical demonstration
of optical diode action in FPPCs using time-domain simulations is
provided. Finally, Section~\ref{sec:SUMMARY} summarizes the paper.

\section{Perfect transmission resonances \protect \\
in multilayers\label{sec:THEORY}}

We consider binary multilayer structures built of two kinds of dielectric
layers (labeled A and B) with thicknesses $d_{\text{A}}$, $d_{\text{B}}$,
linear refractive indices $n_{\text{A}}$, $n_{\text{B}}$, and Kerr
nonlinear coefficients $\chi_{\text{A}}$, $\chi_{\text{B}}$, respectively,
located in a homogeneous dielectric medium with $n=n_{0}$. The building
blocks A and B are assumed to be of the same optical thickness in
the linear regime, so that\begin{equation}
n_{\text{A}}d_{\text{A}}=n_{\text{B}}d_{\text{B}}=\pi c/(2\omega_{0})=\lambda_{0}/4.\label{eq:qwave}\end{equation}
This equality is conventionally called the quarter-wave (QW) condition.
It assures that the transmission $T(\omega)$ and reflection $R(\omega)$
spectra for an arbitrary arrangement of A and B layers is periodic
in frequency with period $2\omega_{0}$. An additional mirror symmetry
is present in each period, so that \cite{our2008,ourPRA10}\begin{equation}
\begin{gathered}T(\omega)=T(\omega+2\omega_{0}),\quad T(\omega_{0}-\Delta\omega)=T(\omega_{0}+\Delta\omega),\\
R(\omega)=R(\omega+2\omega_{0}),\quad R(\omega_{0}-\Delta\omega)=R(\omega_{0}+\Delta\omega).\end{gathered}
\label{eq:symmetries}\end{equation}
Thus all the spectral properties of a given QW multilayer structure,
determined by its layer arrangement, are contained in the region $[0;\omega_{0}]$.
Note that QW multilayers always have a PTR at even multiples of $\omega_{0}$,
so that $|T(2m\omega_{0})|=1$ exactly. 

\begin{figure}
\includegraphics[width=1\columnwidth]{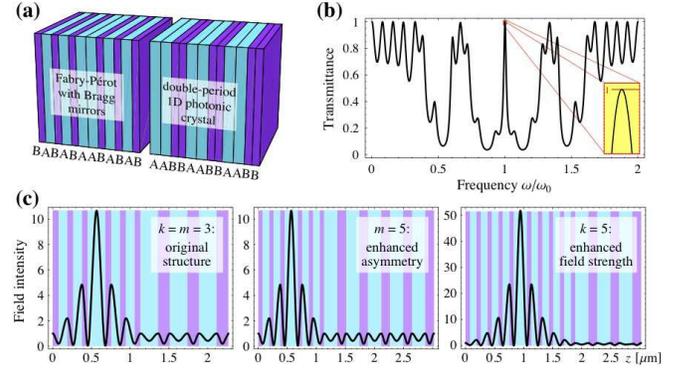}

\caption{(Color online) (a) Design and (b) transmission spectrum of an FPPC
structure $\text{S}^{(3,3)}$ as in Eq.~\eqref{eq:design}, showing
a PTR at $\omega_{0}$ in the inset. (c) Asymmetric field distribution
in structures $\text{S}^{(3,3)}$, $\text{S}^{(3,5)}$ (enhanced spatial
asymmetry), and $\text{S}^{(5,3)}$ (enhanced localization strength)
\cite{ourPRA10}. The materials are chosen as in Ref.~\cite{FabioNewJP10}
to be polydiacetylene (9-BCMU, $n_{\text{A}}=1.55$, $\chi_{\text{A}}=2.5\times10^{-5}\,\text{cm}^{2}/\text{MW}$)
and rutile ($\text{TiO}_{2}$, $n_{B}=2.3$, $\chi_{B}\approx0$).
The thicknesses $d_{\text{A}}$ and $d_{\text{B}}$ are chosen according
to Eq.~\eqref{eq:qwave}. \label{fig:FIG1-intro}}

\end{figure}
When two arbitrary multilayers $\text{\ensuremath{\sa}}$ and $\sb$
are stacked together to form a combined structure $\sa\sb$, it is
easy to relate its spectral properties to those of its constituents
via Airy-like formulas at every frequency \begin{equation}
R_{\sa\sb}=R_{\sa}+\frac{T_{\sa}R_{\sb}T_{\sab}}{1-R_{\sab}R_{\sb}},\quad T_{\sa\sb}=\frac{T_{\sa}T_{\sb}}{1-R_{\sab}R_{\sb}}.\label{eq:Airy_stack}\end{equation}
Here, $\sab$ denotes the inversion of $\sa$, e.g., $\sab=\text{BAABA}$
for $\sa=\text{ABAAB}$. Eqs.~\eqref{eq:Airy_stack} can be used
in a recurrent fashion with Fresnel formulas at dielectric interfaces
acting as the initial conditions, giving rise to an analytical way
of calculating $R(\omega)$ and $T(\omega)$ for any multilayer structure
\cite{ourPRA10}. 

It can be seen from Eqs.~\eqref{eq:Airy_stack} that $R_{\sa}(\omega)=R_{\sb}(\omega)=0$
results in $R_{\sa\sb}(\omega)=0$ and $|T_{\sa\sb}(\omega)|=1$.
That is, frequency-matched PTRs in the constituent structures always
result in a PTR at the same frequency in the combined structure. It
is known that both $\sa=(\text{BA})^{k}(\text{AB})^{k}$ (a Fabry-Pérot
interferometer with a half-wave defect) and $\sb=(\text{AABB})^{m}$
(a 1D PhC with doubled period) have a PTR exactly at $\omega=\omega_{0}$.
As a result, the combined FPPC structure\begin{equation}
\text{S}^{(k,m)}\equiv\sa\sb=(\text{BA})^{k}(\text{AB})^{k}(\text{AABB})^{m}\label{eq:design}\end{equation}
also has a PTR at $\omega=\omega_{0}$ \cite{ourPRA10}. As the light
is strongly localized in $\sa$ (the Fabry-Pérot part of the structure),
the spatial field distribution in the resonant mode is highly asymmetric.
Moreover, larger $k$ enhances the strength of field localization,
whereas larger $m$ increases its spatial asymmetry (Fig.~\ref{fig:FIG1-intro}c),
so they can both be controlled independently and straightforwardly.
This makes the FPPC structures of Eq.~\eqref{eq:design} particularly
attractive for a systematic study of unidirectional transmission involving
a single-cavity resonant mode. In what follows, we are going to analyze
the nonlinear optical properties of $\text{S}^{(n,k)}$ at its PTR
around $\omega_{0}$.

\begin{figure}
\includegraphics[width=1\columnwidth]{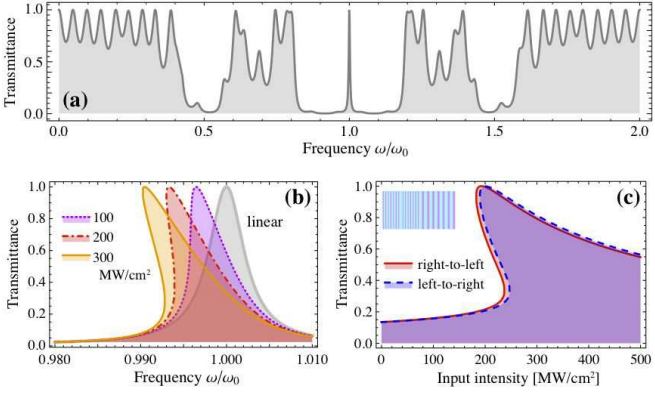}

\caption{(Color online) (a) Linear transmission spectrum, (b) nonlinear transmission
spectrum around $\omega_{0}$ showing PTR bending, and (c) input/output
transmission hysteresis at $\omega=0.9935\omega_{0}$ for the structure
$\text{S}^{(5,5)}$. \label{fig:FIG1a-intro-NL}}

\end{figure}

\section{Nonlinear transmission spectra \protect \\
of asymmetric multilayers\label{sec:PARAMETERS}}

To investigate the optical properties of FPPC multilayers in presence
of Kerr nonlinearity, we assume that the refractive index varies with
field intensity $I$ as \begin{equation}
n_{\text{A,B}}^{(\text{NL})}(z,t)=n_{\text{A,B}}\left(1+\chi_{\text{A,B}}I(z,t)/2\right)\label{eq:Kerr}\end{equation}
where $I=I(z,t)$ varies both in space and in time, making the transmission
spectrum intensity-dependent and potentially multistable \cite{NLTransMat92,diodeNLScalora94,LinLanChinese05}.
To calculate it, we use the standard nonlinear transfer matrix method
with layer subdivision \cite{NLTransMat92} (note that more advanced
methods have been introduced recently \cite{FabioNewJP10}). The example
transmission spectra of $\text{S}^{(5,5)}$ are shown in Fig.~\ref{fig:FIG1a-intro-NL}.
The PTR is seen to undergo characteristic resonance bending, in accordance
with numerous earlier findings \cite{NLTransMat92,LinLanChinese05,LinLanJOSA06,FabioJAP08,FabioNewJP10,FabioTaconaPNFA10}.
Obviously, the resonance bends more strongly as the input intensity
increases, and transmission spectra eventually turn bistable (Fig.~\ref{fig:FIG1a-intro-NL}b)
with hysteresis behavior clearly visible on an output vs.~input intensity
diagram (Fig.~\ref{fig:FIG1a-intro-NL}c). Note that a slight dependence
on the direction of incidence can already be noticed in the hysteresis
loop, although in $\text{S}^{(5,5)}$ the effect is too weak and requires
very high input intensities.

\begin{figure}
\includegraphics[width=1\columnwidth]{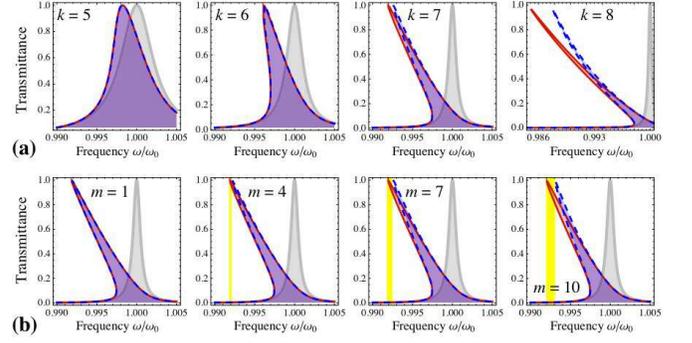}\caption{(Color online) Nonlinear transmission spectra of (a) $\text{S}^{(k,7)}$
for $k=5,6,7,8$ and (b) $\text{S}^{(7,m)}$ for $m=1,4,7,10$. Red
solid and blue dashed lines correspond to right-to-left and left-to-right
propagation of light with input intensity 50 $\mwcm$. Light gray
upright peak is the linear transmission spectrum. Regions of prospective
unidirectional transmission are highlighted as light yellow shaded
areas.\label{fig:FIG2-bending_mk}}

\end{figure}

From the FPPC design principles it is apparent how the changes in
$k$ and $m$ should reflect themselves in nonlinear transmission
spectra. Increasing $k$ will increase both the resonant mode $Q$-factor
(making the resonance peak narrower) and the field enhancement factor
(resulting in a stronger nonlinear interaction and hence a stronger
resonance bending for the same input field intensity). Increasing
$m$ should enhance the asymmetry in the field distribution, so that
the nonlinear transmission of the structure becomes more dependent
on the direction on incidence. This is confirmed in Fig.~\ref{fig:FIG2-bending_mk}.
For moderate values of input intensity ($\leq50\,\mwcm$) asymmetric
transmission starts to manifest itself for $k\geq7$, becoming more
pronounced with increasing $m$. It can be seen that owing to direction-dependent
asymmetry in resonance bending, there is a frequency range where a
high-transmission state only exists for one direction of incidence
(see Fig.~\ref{fig:FIG2-bending_mk}b). In this range, high-quality
unidirectional transmission is expected. 

\begin{figure}[b]
\includegraphics[width=1\columnwidth]{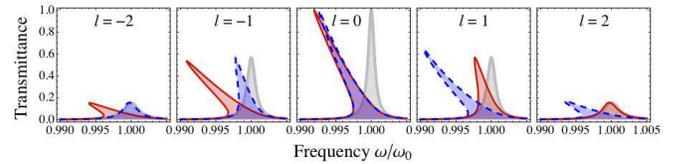}

\caption{(Color online) Nonlinear transmission spectra of the modified structure
as of Eq.~\eqref{eq:design1} for $k=7$, $m=7$, and $l=-2\ldots+2$.
Red solid and blue dashed lines correspond to to right-to-left and
left-to-right propagation of light with input intensity 50 $\mwcm$.\label{fig:FIG3-bending_l}}

\end{figure}

Note in Fig.~\ref{fig:FIG2-bending_mk} that the maximum transmittance
($T_{\text{max}}$) in a bent PTR remains very close to unity despite
the fact that Eq.~\eqref{eq:Kerr} perturbs the QW condition \eqref{eq:qwave}
for $\chi\neq0$. Hence the PTR condition (or, in fact, any recipe
to provide a resonance with near-unity transmittance) is important
to ensure that the reflection losses of an optical diode will remain
low. To demonstrate that the diode performance is very sensitive to
the structure design, we compare $\text{S}^{(7,7)}$ in Eq.~\eqref{eq:design}
with its modification created by shifting the defect in its Fabry-Pérot
part, which can be expressed as \begin{equation}
(\text{BA})^{k-l}(\text{AB})^{k+l}(\text{AABB})^{m}.\label{eq:design1}\end{equation}
\begin{figure}
\includegraphics[width=1\columnwidth]{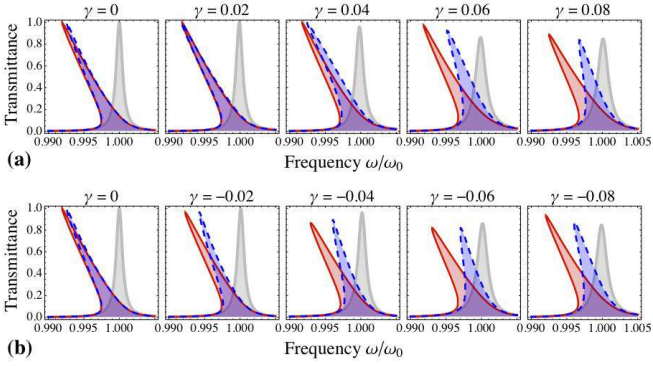}\caption{(Color online) Nonlinear transmission spectra of the modified structure
$\text{S}'^{(7,7)}$ as of Eq.~\eqref{eq:design2} for (a) positive
and (b) negative $\gamma$ ranging between -0.08 and 0.08. Red solid
and blue dashed lines correspond to to right-to-left and left-to-right
propagation of light with input intensity 50 $\mwcm$.\label{fig:FIG4-bending_gamma}}

\end{figure}
The results are shown in Fig.~\ref{fig:FIG3-bending_l}. It is clearly
seen that for the shifted defect ($l\neq0$) the resonance is no longer
a PTR and the linear transmittance comes nowhere near unity, resulting
in poor $T_{\text{max}}$ in the nonlinear regime as well. On the
other hand, it can be seen that the asymmetry in transmission gets
drastically enhanced. So, while the optical diode action severely
suffers in terms of reflection losses, unidirectional transmission
becomes far easier to achieve. The reason is that increasing the number
of periods in a Bragg mirror greatly increases its quality, in turn
increasing the amount of time it takes the light to tunnel through
such mirror into the Fabry-Pérot defect mode. Hence, the difference
of two periods in the $(\text{BA})^{k-1}(\text{AB})^{k+1}$ structure
not only destroys the PTR condition but also creates a much stronger
asymmetric response than that introduced by the $(\text{AABB})^{m}$
part. This is understandable because $(\text{AABB})^{m}$ is free
of internal reflections at $\omega_{0}$, in contrast to $(\text{BA})^{2l}$. 

\begin{figure}[b]
\hfill{}\includegraphics[width=0.75\columnwidth]{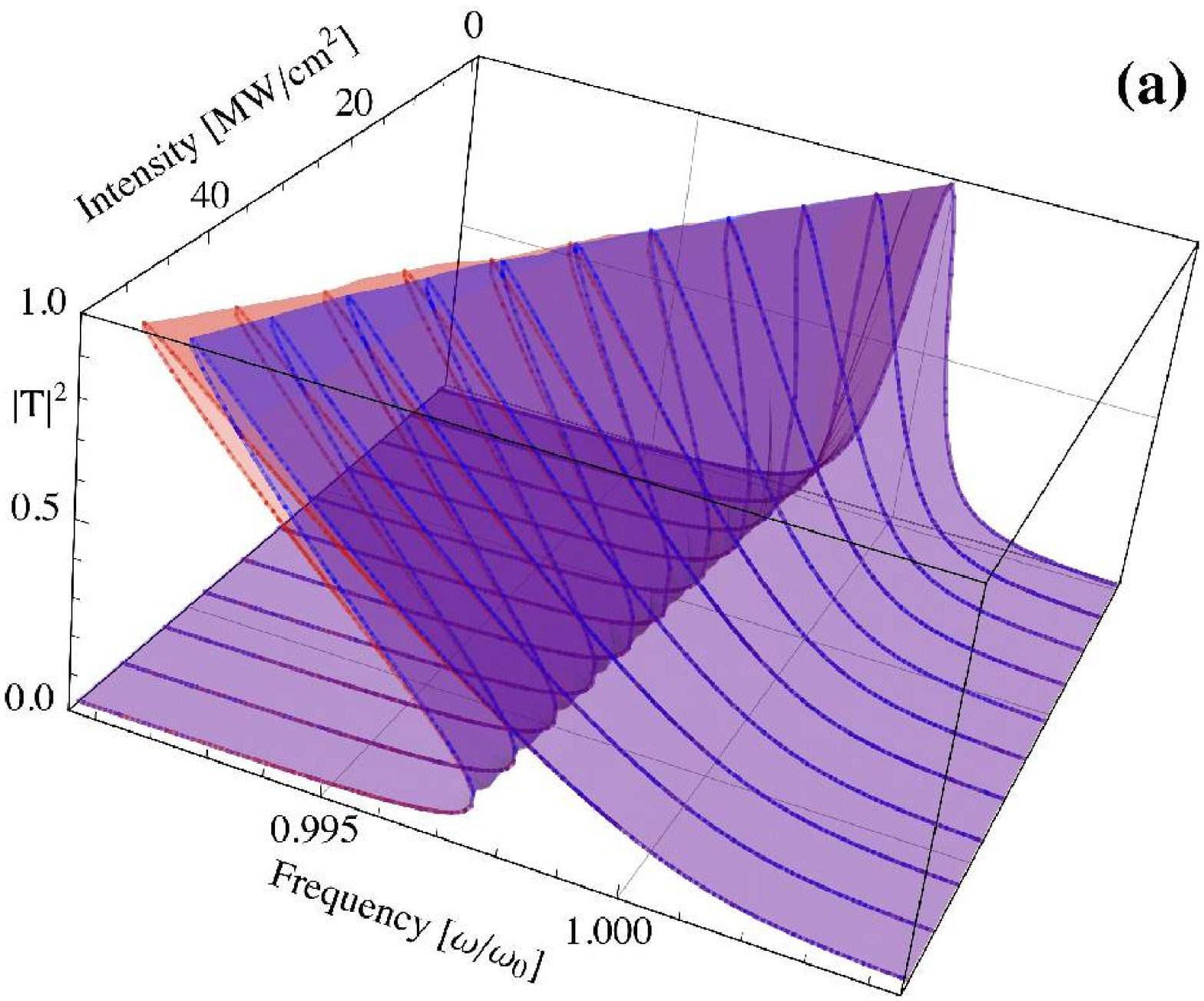}\hfill{}

\hfill{}\includegraphics[width=0.75\columnwidth]{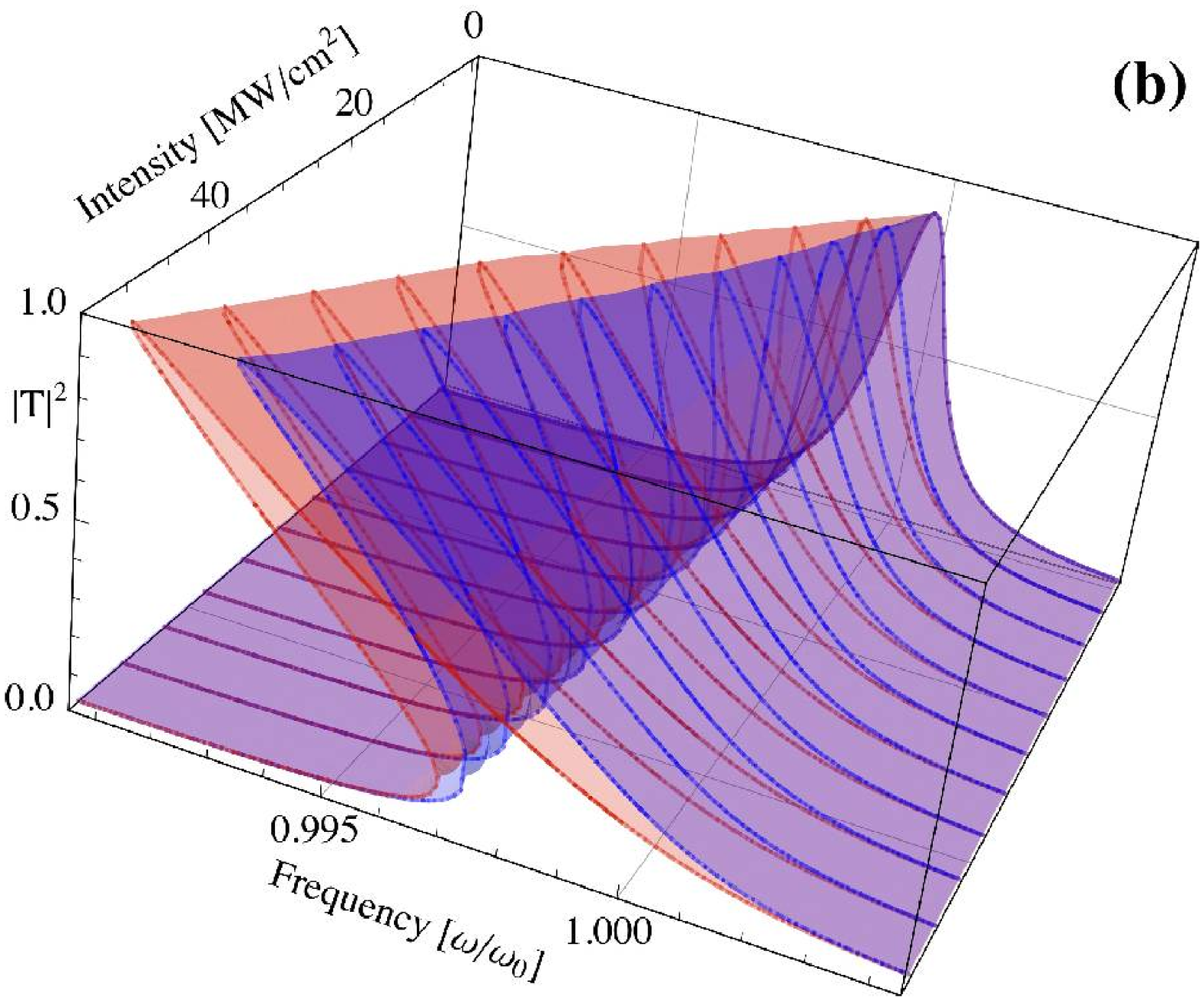}\hfill{}

\caption{(Color online) Nonlinear transmission spectra of (a) $\text{S}^{(7,7)}$
and (b) $\text{S}'^{(7,7)}$ with $\gamma=0.05$, for input intensities
up to 55 $\mwcm$. Two semi-transparent surfaces correspond to two
different directions of light propagation.\label{fig:FIG5-spectra_3D}}

\end{figure}

Therefore, it can be concluded that FPPCs feature a tradeoff between
the maximum transmittance and unidirectionality. Hence it can be expected
that an optimum performance would be achieved with a slight perturbation
of the PTR condition. Consider a structure \begin{equation}
\text{S}'^{(k,m)}=(\text{BA})^{k}(\text{AB})^{k}(\text{A}'\text{A}'\text{BB})^{m}\label{eq:design2}\end{equation}
where $\text{A}'$ denote the A-layers with slightly changed optical
thickness in the photonic crystal part, e.g., by \begin{equation}
d_{\text{A}'}=(1+\gamma)d_{\text{A}}.\label{eq:perturbation}\end{equation}
For small $\gamma$, Fig.~\ref{fig:FIG4-bending_gamma} shows that
the enhancement of unidirectionality more than makes up for a slight
decrease in resonance quality. What is more, for $\gamma\simeq0.04$
one can see that nonlinear $T_{\text{max}}$ becomes \emph{closer}
to unity compared to the linear regime, in the same way as it is seen
for $l=1$ in Fig.~\ref{fig:FIG3-bending_l}. Hence, the $\text{S}'^{(k,m)}$
design not only combines the benefits of structures given by Eqs.~\eqref{eq:design}
and \eqref{eq:design1} but also \emph{restores} the PTR condition
for a properly chosen $\gamma$. The reason is that the slight perturbation
of the QW condition \eqref{eq:qwave} by $\gamma$ is compensated
for by the nonlinear refractive index shift \eqref{eq:Kerr} for a
certain value of input field intensity. Fig.~\ref{fig:FIG5-spectra_3D},
comparing the 3D nonlinear transmission spectra of $\text{S}'^{(7,7)}$
vs.~$\text{S}^{(7,7)}$, illustrates that the asymmetry is much more
pronounced for the former case, while $T_{\text{max}}$ remains close
to unity in both structures. With the effect of PTR restoration taken
into account, the $\text{S}'^{(7,7)}$ design is unambiguously superior
in terms of the optical diode performance. %

\section{Time-domain simulations \protect \\
of optical diode action\label{sec:FDTD}}

\begin{figure}
\includegraphics[width=1\columnwidth]{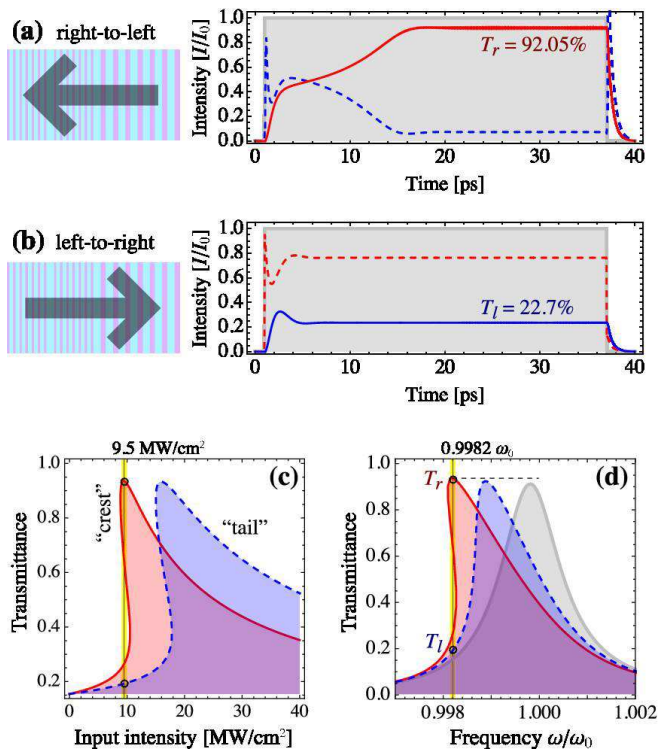}

\caption{(Color online) Results of FDTD simulations for passive diode action
in a $\text{S}'^{(7,7)}$ structure for $I_{0}=9.5\,\mwcm$ and $\omega=0.9982\omega_{0}$:
(a) with incident wave from the right and (b) with incident wave from
the left. Solid lines, transmittance; dashed lines, reflectance. Blue
(darker) color, left-to-right propagation; red (lighter) color, right-to-left
propagation. In the background, quasi-monochromatic incident pulse
is shown in light gray. The analytically derived operating point of
the diode in (c) intensity hysteresis loop and (d) nonlinear transmission
spectra is also presented, in the same form as Figs.~\ref{fig:FIG2-bending_mk}--\ref{fig:FIG4-bending_gamma}.\label{fig:FIG6-FDTD}}

\end{figure}

In this section, we demonstrate the unidirectional transmission and
optical diode action in optimized FPPC structures as given by Eq.~\eqref{eq:design2}
by means of direct time-domain numerical simulations. We use the 1D
Lax-Wendroff method \cite{LaxWendroff} as implemented in Refs.~\cite{Smirnov2000,Smirnov2004},
which is analogous to a commonly known Yee's finite-difference time-domain
scheme. 

In accordance with the guidelines of the previous sections, $\text{S}'^{(7,7)}$
with $\gamma=0.05$ is examined as a model structure. As seen in Fig.~\ref{fig:FIG5-spectra_3D}b,
this structure features a maximum linear transmission of 91.2\%. In
the nonlinear regime, its right-to-left transmittance $T_{r}$ increases
towards unity due to the PTR restoration effect described above, while
left-to-right transmittance $T_{l}$ at the same frequency is low.
The operating input intensity was chosen at a rather moderate value
of $9.5\,\mwcm$ in order to be able to work in the simple, passive
regime. Higher input intensities ($>20\,\mwcm$), where the resonance
peak gets sufficiently bent, make it increasingly difficult to excite
the upper branch of the hysteresis loop by means of the incident wave
alone. This can be overcome by using pump-assisted operation \cite{LinLanChinese05,LinLanOE08}
or modulated incident wave \cite{FabioNewJP10,FabioTaconaPNFA10},
which is rather unfavorable for optical diode applications but may
be promising for all-optical modulator designs.

The structure was discretized in the $z$ direction with spatial mesh
size $\Delta z=5\,\text{nm}$, which corresponds to 30--50 cells per
layer. The time step was determined by the condition that the Courant
number $\nu=1$, so $\Delta t=0.01667\,\text{fs}.$ Such parameters
ensure that the effects of numerical dispersion can be neglected in
our 1D calculations. Indeed, the test runs without nonlinearity confirm
that the resonant frequency $\omega_{0}=2\pi\times12.161\times10^{14}$~Hz
matches the results of linear transfer matrix calculations.

Unidirectional transmission is demonstrated by examining the steady-state
response of the nonlinear multilayer to a continuous wave incident
from either side of the structure. To simulate this, we use a quasi-monochromatic
wave packet with a rectangular envelope, long enough for the steady-state
response to develop. The carrier frequency, determined by the input
intensity values, was chosen at $\omega=0.9982\omega_{0}$ or $12.1388\times10^{14}$
Hz. To avoid numerical artifacts caused by abrupt onsets of the rectangular
envelope, Gaussian-like transient excitations with decay time of 0.03~ps
were applied immediately before turning on and after turning off the
rectangular quasi-monochromatic pulse (at 1 and 37 ps, respectively). 

The results are presented in Fig.~\ref{fig:FIG6-FDTD}a--b. It is
seen that right-to-left transmittance $T_{r}$ eventually reaches
92.05\%, slightly exceeding the maximum linear transmittance of the
structure (91.2\%), although the transient processes are seen to last
for up to 15~ps. In the reverse direction, the transmittance $T_{l}$
remains below 23\%. The on-off contrast can be improved by increasing
both the operating intensity and $\omega$, which will also decrease
the transient time in the right-to-left propagation, but this will
decrease $T_{r}$. The maximum nonlinear correction to the refractive
index, monitored in the middle of the Fabry-Pérot cavity during entire
simulation, does not exceed the values of 0.0085 (right-to-left propagation)
and 0.003 (left-to-right propagation). These values are even lower
than the estimation of 0.01 made in Ref.~\cite{FabioNewJP10}, so
the structure is expected to remain stable for the chosen intensities.

For comparison, the operating regime of the proposed optical diode
derived from the nonlinear transfer matrix method is presented in
Fig.~\ref{fig:FIG6-FDTD}c--d. Analytical calculation is seen to
yield slightly better performance figures ($T_{r}=93.5\%$, $T_{l}=19.5\%$).
These small discrepancies are attributed to spatial discretization
present in both time-domain and nonlinear transfer matrix methods
and can be seen as the effect of $\sim1-2\%$ structural disorder.
Other than that, analytical and numerical results show a very good
agreement. 

It is remarkable that the passive unidirectional transmission is observed
at the lower-intensity {}``crest'' of the hysteresis loop (see Fig.~\ref{fig:FIG6-FDTD}c)
rather than at its higher-intensity ``tail'' as in earlier works
\cite{LinLanChinese05}. Previously, only active or pump-assisted
optical diode action was achieved in this regime \cite{LinLanOE08,FabioNewJP10,FabioTaconaPNFA10}.

Note, finally, that the demonstrated optical diode action occurs in
the same input intensity range as in the recently reported Thue-Morse
multilayers \cite{FabioNewJP10,FabioTaconaPNFA10} but with the use
of a structure less than half as thin and containing half as many
layers (56 versus 128). This results from an increased field localization
strength in FPPC structures. Another advantage of the FPPC design
is that it allows to purposely engineer a PTR-featuring resonant mode
with desired properties, rather than rely on naturally occurring PTRs
in photonic quasicrystals where not all resonant modes are suitable
for unidirectional operation (see \cite{FabioNewJP10}). The operating
intensities can be brought further down by increasing $k$, with a
relatively insignificant incease in the number of layers. However,
this will also increase the length of the transient response of the
structure, effectively resulting in a tradeoff between energy efficiency
and maximum attainable bandwidth when such an optical diode is used
in communication systems.

\section{Conclusions and outlook\label{sec:SUMMARY}}

In summary, we have theoretically investigated the optical properties
of asymmetric Kerr-nonlinear multilayers with perfect transmission
resonances. Using the nonlinear transfer matrix method, it is found
that FPPC structures of the type $(\text{BA})^{k}(\text{AB})^{k}(\text{AABB})^{m}$
exhibit both pronounced unidirectionality and high transmission, which
makes these structures suitable for nonlinear optical diode action
in the passive regime with low reflection losses. A trade-off is found
between the maximum transmission and unidirectionality, subject to
tuning by slightly perturbing the PTR condition. An effect of PTR
restoration is observed in perturbed FPPC structures. Theoretical
predictions are confirmed in direct time-domain numerical simulations,
yielding more than 92\% transmission for the input intensity of <10
$\mwcm$. This value exceeds the linear maximum transmittance of 91.2\%,
further confirming the PTR restoration effect. Passive optical diode
regime was demonstrated where only pump-assisted or active operation
was reported previously \cite{LinLanChinese05,FabioNewJP10,FabioTaconaPNFA10},
and the proposed geometry offers the same performance as recently
reported Thue-Morse multilayers \cite{FabioNewJP10,FabioTaconaPNFA10}
but with less than half as many layers. 

The results obtained can also be viewed as a systematic investigation
of a single photonic-multilayer resonant cavity in the nonlinear regime.
They can be used as a starting point for investigating more sophisticated
multilayer geometries combining spatial asymmetry and high transmission.
For example, recent studies report that coupled-cavity resonances
in multiple-defect PhCs \cite{Smirnov2002} and photonic quasicrystals
\cite{FabioNewJP10} have very different character of nonlinear bending
and different properties of unidirectional transmission, compared
to single-cavity resonances. To examine PTRs created by multiple Fabry-Pérot
elements coupled in an asymmetric fashion would therefore be a very
interesting extension of the principles derived in the present work. 
\begin{acknowledgments}
The authors are very grateful to Victor Grigoriev for inspiring discussions
and helpful suggestions. One of us (S.V.Z.) wishes to acknowledge
partial support by the Natural Sciences and Engineering Research Council
of Canada (NSERC).\end{acknowledgments}

\end{document}